\documentstyle[preprint,eqsecnum,aps]{revtex}
\begin{document}
\setlength{\unitlength}{1mm}
\newcommand{\I}{\mbox{\rm I} \hspace{-0.5em} \mbox{\rm I}\,}
\newcommand{\te}{\theta}
\newcommand{\be}{\begin{equation}}
\newcommand{\ee}{\end{equation}}
\newcommand{\bea}{\begin{eqnarray}}
\newcommand{\ena}{\end{eqnarray}}
\newcommand{\tra}{\triangle\theta_}
\newcommand{\emt}{\varepsilon}
\newcommand{\ham}{\hat{{\cal H}}}
\newcommand{\inv}{\hat{{\cal I}}}

\title{{\hfill} {\sl Submitted} \\ 
Landau damping of partially incoherent
Langmuir waves}

\author{R.~Fedele$^1$, P.K.~Shukla$^2$, M. Onorato$^3$,
D.~Anderson$^4$, and M.~Lisak$^4$}

\address{$^1$ \it Dipartimento di Scienze Fisiche, Universit\`{a} di
Napoli ''Federico II'' and INFN Sezione di Napoli,
Complesso Universitario di M.S. Angelo,\\ Via Cintia,
I-80126 Napoli, Italy}
\address{$^2$ \it Institut f\"{u}r Theoretische Physik IV,
Ruhr-Universit\"{a}t Bochum \\ D-4470 Bochum, Germany}
\address{$^3$ \it Dipartimento di Fisica Generale, Universit\`{a} di
Torino, \\Via Pietro Giuria 1, 10125 Torino, Italy}
\address{$^4$ \it Department of Electromagnetics, Chalmers
University of Technology, SE-412 96, G\"{o}teborg, Sweden}
\maketitle
\begin{abstract}
It is shown that   partial incoherence, in the form of
stochastic phase noise, of a  Langmuir wave in an
unmagnetized plasma  gives rise to a  Landau-type
damping. Starting from the Zakharov equations, which
describe the nonlinear interaction between Langmuir and
ion-acoustic waves,  a  kinetic equation is derived for
the plasmons by introducing the Wigner-Moyal transform of
the complex Langmuir wave field.  This equation is then
used to analyze the stability properties of small
perturbations on a stationary solution consisting of a
constant amplitude wave with stochastic phase noise. The
concomitant dispersion relation exhibits the phenomenon
of Landau-like damping. However, this damping differs
from the classical Landau damping in which a Langmuir
wave, interacting with the plasma electrons, loses
energy. In the present process, the damping is
non-dissipative and is caused by the resonant interaction
between an instantaneously-produced disturbance, due to
the parametric interactions, and a partially incoherent
Langmuir wave, which can be considered as a
quasi-particle composed of an ensemble of partially
incoherent plasmons.
\end{abstract}

\bigskip

\noindent
PACS number(s): 52.35.Mw, 52.35.Fp, 52.25.Dg, 03.40.Kf

\bigskip


\newpage
More than fifty years ago, Landau \cite{9} discovered a
damping of electron plasma waves whose linear dynamics is governed
by the Vlasov-Poisson system of equations. This damping was caused
by resonant wave-electron interactions. The Landau theory clearly
shows that the decay rate of  the wave energy due to the
interaction with the resonant electrons of the system (satisfying
$p=\omega/k$, where $p$ denotes the electron velocity and $\omega$
and $k$ are the frequency and wave number of the electron plasma
wave ) is proportional to the first derivative of the equilibrium
distribution function $\rho_0 (p)$ of the electrons. Typically the
shape of $\rho_0 (p)$ is such that $d\rho_0 (p=\omega/k)/dp<0$,
which implies that  there are more particles with $p < \omega/k$
(which gain energy from the wave)  than  with $p
>\omega/k$ (which give energy to the wave). Consequently, the net
result is that  the wave is  damped.  However, finite
amplitude Langmuir waves in plasmas can be created when
some free energy sources, such as electron and laser
beams, are available in the system. In the past, several
authors \cite{nishikawa}-\cite{Zakmushrub} have considered
the nonlinear coupling between high-frequency Langmuir
and low-frequency ion-acoustic waves. Physically, this
coupling occurs because a large amplitude plasma wave,
interacting with small density ripples, produces a
current which becomes the source for an envelope of
Langmuir waves. The low-frequency ponderomotive force of
the latter reinforces the density modulation in such a
way as to produce a space-charge electric field, which
varies on a time scale longer than the electron plasma
period. Thus, within the slowly-varying envelope
approximation, the dynamics of the Langmuir wavepacket is
governed by a nonlinear Schr\"{o}dinger equation (NLSE),
while the non-resonant density ripples follow an
ion-acoustic wave equation driven by a ponderomotive
force created by the Langmuir wave). In his classic
paper, Zakharov \cite{Zakharov} demonstrated that these
nonlinearly coupled equations admit a class of
modulationally unstable stationary solutions and also
exhibit the phenomenon of wave collapse.\\  In this
paper, we present a  Landau damping of partially
coherent Langmuir waves whose dynamics is governed by the
Zakharov equations. In fact, we show that the latter can
be reduced to a kinetic-like equation for a
Wigner-Moyal-like transform for the amplitude of the
electric field associated with the Langmuir wave. Based
on this kinetic equation we consider the dynamics of
small perturbations on a  partially coherent background
solution consisting of a constant amplitude and a
stochastically varying phase. The concomitant  dispersion
relation depends strongly  on the spectral energy density
of the Langmuir wave and  a Landau-type damping of the
perturbations is found due to the resonant interaction
between the Langmuir quasi-particle and the
instantaneously created ripples.

As it is well known, the nonlinear propagation of a
one--dimensional Langmuir wave--packet in an unmagnetized
plasma with unperturbed density $n_0$, is governed by the
Zakharov equations \cite{Zakharov}
\begin{equation}
i{\partial E\over\partial t}~+~{3 v_{te}^{2}\over
2\omega_{pe}}{\partial^2 E\over\partial
x^2}~-~\omega_{pe}{\delta n\over 2n_0}E~=~0~~~,
\label{zakharov-1}
\end{equation}
\begin{equation}
\left({\partial^2\over\partial t^2}~-~c_{s}^{2}{\partial^2\over\partial x^2}
\right){\delta n\over 2n_0}~=~-{c_{s}^{2}\over 2}{\partial^2\over\partial x^2}
\left({|E|^2 -|E_0|^2\over 4\pi n_0 T_e}\right)~~~,
\label{zakharov-2}
\end{equation}
where $E$ is the slowly varying complex electric field
amplitude, $\delta n$ is the density fluctuation,
$v_{te}$ is the electron thermal  velocity, $\omega_{pe}$
is the electron plasma frequency, $T_e$ is the electron
temperature, $c_{s}$ is the sound speed, and $E_0$ is the
unperturbed electric  field amplitude. Moreover, $x$ and
$t$ play the role of configurational space coordinate
(longitudinal coordinate with respect to the wavepacket
centre) and time, respectively.   We will show that the
nonlinear propagation of a Langmuir wavepacket, governed
by the above Zakharov system, can be suitably described
in phase space by a kinetic--like equation within a
framework similar to the one of the {\it Vlasov-Poisson}
system. A similar approach has recently been  used  to
study the modulational instability  of a large-amplitude
electromagnetic  wavepacket as well as the coherent
instability analysis of intense charged--particle  beam
dynamics  \cite{5,6}. Within this framework, we show that
a phenomenon similar to the classical Landau damping
occurs for partially coherent  plasma waves. In
particular, this phenomenon tends to suppress
 the modulational instability. To this end, the
above system of equations (\ref{zakharov-1}) and
(\ref{zakharov-2}) can be cast in the  following form
\begin{equation}
i\alpha{\partial \Psi\over\partial s}~+~{\alpha^{2}\over
2}{\partial^2
\Psi\over\partial
x^2}~-~U~\Psi~=~0~~~,
\label{zakharov-1-bis}
\end{equation}
\begin{equation}
\left({\partial^2\over\partial s^2}~-~\mu^{2}{\partial^2\over\partial x^2}
\right)U~=~-{\partial^2\over\partial x^2}
\left(\langle|\Psi|^2 -|\Psi_0|^2\rangle\right)~~~,
\label{zakharov-2-bis}
\end{equation}
where the angle brackets account for the statistical
ensemble average {\it \'{a}la} Klimomtovich \cite{a} due to
the partial incoherence of the waves,
$s=\sqrt{3}\lambda_{De}\omega_{pe}t$ ($\lambda_{De}\equiv
v_{te}/\omega_{pe}$ is the electron Debye length),
$\alpha
=\sqrt{3}\lambda_{De}$, $U=\delta n / 2n_0$,
$\mu=\sqrt{m_{e}/3m_{i}}$  ($m_{e}$ and $m_{i}$ are the electron
and the ion masses, respectively), and
\begin{equation}
\Psi~=~{\mu E\over \sqrt{8\pi n_0 T_e}}~~~,
\label{Psi}
\end{equation}
\begin{equation}
\Psi_0~=~{\mu E_0\over \sqrt{8\pi n_0 T_e}}~~~.
\label{Psi-zero}
\end{equation}
Note that (\ref{Psi-zero}) can be written as
\begin{equation}
\rho_0
\equiv |\Psi_0|^2=\mu^2{\epsilon\over\epsilon_{T}}~~~,
\label{Psi-zero-bis}
\end{equation}
where $\epsilon\equiv E_{0}^{2}/ 8\pi$ and
$\epsilon_T\equiv n_0 T_e$ are the electric energy
density and the electron thermal energy density,
respectively. According to the theory modelled by the
Zakharov system (\ref{zakharov-1-bis}) and
(\ref{zakharov-2-bis}), $\epsilon /\epsilon_T >>1$ (i.e,
the ponderomotive effect is larger than the thermal
pressure effect). Note also that Eq.
(\ref{zakharov-2-bis}) implies that $U$ is a functional
of $|\Psi|^2$, namely $U=U\left[|\Psi|^2\right]$.
Consequently, once (\ref{zakharov-2-bis}) is combined
with (\ref{zakharov-1-bis}), the latter becomes a
nonlinear Schr\"{o}dinger equation (NLSE)  with the nonlinear
term $U\left[|\Psi|^2\right]$. This implies that,
as in Quantum Mechanics, one can introduce the
density matrix
 \cite{a} as a sort of "two-points
correlation function" and then the Wigner-Moyal transform
\cite{b} which is solution of the von Neumann-Weyl
equation \cite{c}. In case of surface gravity waves, whose dynamics and
instability is described in deep water
by a nonlinear Schr\"{o}dinger equation,
such a kind of approach has been given in the pioneering works
by Alber \cite{e}, Crawford et al. \cite{f} and Janssen
\cite{g}. More recently, a similar
approach has been developed for the propagation of
electromagnetic wavepackets in nonlinear media where an
appropriate kinetic equation is able to show a random
version of the modulational instability \cite{5,6,j}. In
particular, this approach can be also seen as an
application of the Klimontovich statistical average
method \cite{j}. The transition to the phase-space $x-p$
($p\equiv dx/ds$ is the conjugate variable of $x$) allows
us to write a von Neumann equation \cite{a}  for
$w(x,p,s)$. In fact, although, we are dealing with a
classical system, by following Alber \cite{e}, one can
introduce the correlation function (whose
corresponding meaning in Quantum Mechanics is nothing but
the density matrix for mixed states)
\begin{equation}
\rho(x,x',s)= \langle\Psi^*(x,s) \Psi (x',s)\rangle~~~,
\label{wigner}
\end{equation}
where the statistical ensemble average takes into account
the incoherency of the waves. The technique was
successfully introduced in the field of statistical
quantum mechanics to describe the dynamics of a system in
the classical space language \cite{e}-\cite{g}. Thus one
can define the following Wigner-Weyl transform \cite{b}
\begin{equation}
w(x,p,s) = {1 \over 2 \pi \alpha}
\int_{-\infty}^{\infty}
\rho\left(x + { y \over 2},~ x - { y \over
2}, ~s\right)~\exp\left(i{p y\over \alpha} \right)~dy~~~,
\label{12-0}
\end{equation}
which allows us to transit from the configuration space
to the phase space. In fact, $w(x,p,s)$ is governed by
the following kinetic-like equation
(von Neumann-like equation \cite{a}~)
\begin{equation}
{\partial w\over \partial s} + p {\partial w\over
\partial x}
- \sum_{n=0}^{\infty}{(-1)^{n}\over \left(2n+1\right)!}
\left( { \alpha \over 2} \right)^{2 n}
{\partial^{2n+1}U\over\partial x^{2n+1}}
{\partial^{2n+1}w\over\partial p^{2n+1}}~=~0~~~.
\label{a4}
\end{equation}
On the other hand, since (\ref{12-0}) implies that
\begin{equation}
\langle |\Psi|^2\rangle~=~\int_{-\infty}^{\infty}w(x,p,s)~dp~~~,
\label{x-projection}
\end{equation}
the functional $U\left[|\Psi|^2\right]$ can be expressed
 as a functional of $w$, namely $U=U\left[w\right]$.
Consequently, (\ref{zakharov-2-bis}) can be written as
\begin{equation}
\left({\partial^2\over\partial s^2}~-~\mu^{2}{\partial^2\over\partial x^2}
\right)U~=~-{\partial^2\over\partial x^2}
\left(\int_{-\infty}^{\infty}w~dp -\int_{-\infty}^{\infty}w_0~dp\right)~~~,
\label{zakharov-2-ter}
\end{equation}
where $w_0$ is the Wigner-Moyal transform of $\Psi_0$ ($\rho_0
=|\Psi_0|^2 = \int_{-\infty}^{\infty}~w_0~dp$).  The above scheme
allows us   to carry out a stability analysis of the stationary
Langmuir wave by considering the linear dispersion relation of
small amplitude perturbations.
We start from the equilibrium state: $w = w_0 (p)$, $U=U_0=0$, and
perturb the system according to
\begin{equation}
w(x,p,s)=w_0 (p) + w_1 (x,p,s)~~~,
\label{rho-perturbation}
\end{equation}
\begin{equation}
U(x,s)~=~U_0~+~U_1(x,s)~=~U_1(x,s)~~~,
\label{U-perturbation}
\end{equation}
where $w_1 (x,p,s)$ and $U_1 (x,s)$ are first-order
quantities. Consequently, (\ref{a4}) and
(\ref{zakharov-2-ter}) can be linearized as follows
\begin{equation}
{\partial w_1\over \partial s} + p {\partial w_1\over
\partial x}
=~\sum_{n=0}^{\infty}{(-1)^{n}\over \left(2n+1\right)!}
\left( { \alpha \over 2} \right)^{2 n}
{\partial^{2n+1}U_1\over\partial x^{2n+1}}
w_{0}^{(2n+1)}~~~,
\label{a4-2}
\end{equation}
\begin{equation}
\left({\partial^2\over\partial s^2}~-~\mu^{2}{\partial^2\over\partial x^2}
\right)U_1~=~-{\partial^2\over\partial x^2}
\left(\int_{-\infty}^{\infty}w_1~dp\right)~~~,
\label{zakharov-2-linearized}
\end{equation}
where $w_{0}^{(2n+1)}\equiv d^{2n+1}\rho_0/dp^{2n+1}$. By
introducing the Fourier transform of $U_1(x,s)$ and $w_1
(x,p,s)$, i.e.
\begin{equation}
U_1(x,s)=\int_{-\infty}^{\infty}~dk~\int_{-\infty}^{\infty}~d\omega~
\widetilde{U_1}(k ,\omega)~\exp\left(ik x-i\omega
s\right)~~~,
\label{U-1-bis}
\end{equation}
\begin{equation}
w_1(x,p,s)=\int_{-\infty}^{\infty}~dk~\int_{-\infty}^{\infty}~d\omega~
\widetilde{w_1}(k,p,\omega)~\exp\left(ik x-
i\omega s\right)~~~,
\label{w-1}
\end{equation}
and substituting (\ref{U-1-bis}) and (\ref{w-1}) into
Eq.s (\ref{a4-2}) and (\ref{zakharov-2-linearized}),  we
readily obtain the following dispersion relation
\begin{equation}
\omega^2~-~\mu^2 k^2~=~k^2\int_{-\infty}^{\infty}~{w_0
\left(p+\alpha k /2\right)~-~w_0
\left(p-\alpha k /2\right)\over \alpha k}~
{dp\over p-\omega /k}~~~.
\label{dispersion-relation}
\end{equation}

We consider a Lorenzian spectrum of the form:
\begin{equation}
w_0(p)=\frac{\rho_0} {\pi} \frac{\Delta} {\Delta^2+p^2},
\end{equation}
where $\Delta$ is the width of the spectrum. Dispersion
relation (\ref{dispersion-relation}) becomes:
\begin{equation}
\left(\omega^2-\mu^2 k^2\right)\left(\omega^2-\alpha^2 k^4/4 +
2 i k \Delta  \omega-k^2 \Delta^2 \right)
~=~\rho_0 k^4~~~,
\label{lorenzian-dispersion-relation}
\end{equation}
As the simplest case, we take the limit of $\Delta \rightarrow 0$.
This case corresponds to
a coherent monochromatic wave-train and the dispersion relation reduces to
\begin{equation}
\left(\omega^2-\mu^2 k^2\right)\left(\omega^2-\alpha^2 k^4 /4\right)
~=~\rho_0 k^4~~~,
\label{monochromatic-dispersion-relation}
\end{equation}
which predicts the well known result of a purely growing
instability \cite{Zakharov}.

In the following, we concentrate on the limit of small
$\alpha k << 1$ and $\omega /k >>1$. We will show that
during its nonlinear evolution the Langmuir wavepacket
can exhibit a phenomenon, similar to the Landau damping
\cite{Dawson}, between parametrically-driven non-resonant
density ripples and Langmuir quasi-particles. It is easy
to verify that the condition $\alpha k << 1$ (i.e.,
$\lambda_{De} k<<1$) implies that
\begin{equation}
{w_0\left(p+\alpha k /2\right)~-~w_0
\left(p-\alpha k /2\right)\over \alpha k}~\approx
dw_{0}/dp~\equiv w_{0}^{'}~~~.
\label{k-small-approximation}
\end{equation}
Consequently, (\ref{dispersion-relation}) becomes
\begin{equation}
\omega^2~=~k^2~\int~{w_{0}^{'}
\over p-\omega /k}~dp~~~,
\label{k-small-dispersion-relation}
\end{equation}
where, using the assumption $\omega /k >>1$, we can neglect the
term  involving $\mu^2$ because $\mu <<1$.
\newline  Note that (\ref{k-small-dispersion-relation})
is formally identical to the linear dispersion relation
that holds for warm plasma waves \cite{9} or for
non-monochromatic charged-particle bunches in circular
accelerating machines for an arbitrary complex coupling
impedance \cite{7}. Consequently, the dispersion relation
(\ref{k-small-dispersion-relation}) allows us to predict
a sort of  Landau damping \cite{9}, as described in
plasma physics  for linear plasma waves as well as in
charged-particle beam physics. A similar effect, which
has been called {\it quantum-like Landau damping} (QLLD),
has  recently been predicted for the nonlinear
propagation of a large-amplitude electromagnetic waves in
nonlinear media, such as optical fibers and plasmas as
well as for the quantum-like description of
charged-particle beams in an accelerating machine
\cite{5,6,j}. In order to investigate this phenomenon
also for large-amplitude Langmuir waves, we apply the
standard procedure \cite{14} used for the Landau damping
of a linear plasma wave, which involves integrating the
above dispersion relation in the complex plane,
evaluating both the Cauchy principle value (PV)  and the
semi-residue in $p= \omega /k$. The result is
\begin{equation}
\omega^2~=~k^2 \int_{PV}~{w_{0}^{'}
\over p-\omega /k}~dp~+
~i\pi k^2 w_{0}^{'}(\beta)~~~,
\label{landau-dispersion-relation}
\end{equation}
For a symmetric sufficiently smooth stationary background
distribution, $w_0 (p)$, we have
\begin{equation}
\omega^2~=~\left(1+{3\Delta^{2}
\over \rho_{0}}\right)k^2~+
~i\pi k^2 w_{0}^{'}(\beta)~~~,
\label{reactive-dispersion-relation-2}
\end{equation}
where $\Delta\equiv \left(\langle p^2 \rangle\right)^{1/2}$
denotes the r.m.s. width of $w_0 (p)$, and $\beta \equiv
\rho_{0}^{1/4}\left(1+3\Delta^{2} / 4\rho_{0}^{2}\right)$. Note
that, according to the above hypothesis, we have $\beta >>1$ and
$\Delta^2 /\rho_0 <<1$. This implies that $\rho_0 =|\Psi_0|^2
>>1$, and from (\ref{Psi-zero-bis}) it follows that $\epsilon
/\epsilon_T >>1/\mu^2$. Eq. (\ref{reactive-dispersion-relation-2})
clearly shows that the damping rate is proportional to the
derivative of the Wigner distribution $w_0$. This  is formally
similar  to the expression for the Landau damping of a linear
plasma wave in a warm  unmagnetized  plasma. In fact, writing
$\omega$ in the complex form $\omega =\omega_R + i\omega_I$,
substituting it in (\ref{landau-dispersion-relation}) and then
separating the real and   imaginary parts, we obtain
\begin{equation}
\omega_{R}^{2}~-~\omega_{I}^{2}~=~k^2\int_{PV}~{w_{0}^{'}
\over p-\omega /k}~dp~~~,
\label{real-part}
\end{equation}
and
\begin{equation}
\omega_{I}~=~{\pi \over 2} {k^2\over \omega_R}
w_{0}^{'}(\beta)~~~.
\label{imaginary-part}
\end{equation}
The Landau damping rate of the Langmuir wave in the appropriate
unities is $\gamma =\sqrt{3}\omega_{pe}\lambda_{De}\omega_I$. For example,
for a Gaussian wavepacket spectrum, i.e. $w_0 (p)= \left(\rho_0
/\sqrt{2\pi \Delta^{2}}\right)\exp\left(-p^2 /2\Delta^{2}
\right)$, one obtains
\begin{equation}
\gamma = -\left({3\pi\over 8}\right)^{
1/2}{\mu^{5/2}
\over\Delta^3}
\left({\epsilon\over \epsilon_T}\right)^{5/4}
~\omega_{pe}\lambda_{De}k~\exp\left[-{\mu
\over 2\Delta^{2}}\left({\epsilon\over\epsilon_T}
\right)^{1/2}\right]~~~,
\label{imaginary-part-1}
\end{equation}
where  higher-order terms have been neglected. The present damping
mechanism differs from the standard Landau damping in that here
$w_0 (p)$ does not represent the equilibrium velocity distribution
of the plasma electrons, but it can be considered as  the
"kinetic" distribution of all Fourier components of the partially
incoherent large-amplitude Langmuir wave in the warm plasma.
 In terms of plasmons, we realize that $w_0 (p)$ represents
the distribution of the partially incoherent plasmons that are
distributed in $p$-space (i.e. in $k$-space) with a finite
"temperature". Consequently, the QLLD described here is due to the
partial incoherence of the wave which corresponds to a
finite-width Wigner distribution of the plasmons.

In this paper, we have carried out an analysis for a
Landau-type damping in plasmas containing partially
incoherent Langmuir waves. Starting from the Zakharov
system of equations and employing the Wigner-Moyal
quasi-distribution function for the complex Langmuir wave
electric field amplitude,  it has been shown that the
Zakharov system of equations can be converted into a pair
of coupled evolution equations in phase space. For a
monochromatic and coherent wave ($w_0(p) =\rho_0\delta
(p)$), we have obtained a dispersion relation which
exhibits a purely growing instability
 \cite{Zakharov}. Furthermore, for a broad-band
Langmuir wave spectrum, corresponding to a partially incoherent
constant amplitude wave, we have investigated the properties of
small perturbations for $\alpha k<<1$ (i.e., $\lambda_{De} k<<1$),
and very large  dimensionless phase velocity (i.e., $\omega /k
>>1$).   Our dispersion relation for
this case exhibits a sort of weak Landau damping,  very similar to
the one predicted for the Vlasov-Poisson system for linear plasma
waves. The physical origin of  this phenomenon is attributed to
the "non-monochromatic" behaviour of the Wigner spectrum of the
Langmuir  wave, which can be considered as an "ensemble of
incoherent plasmons". Actually, a broad-band spectrum  of
incoherent plasmons, forming the "wave packet", interact
individually with the self-excited density ripples. Similar to the
standard Landau damping,  where the electrons interact
individually with a linear plasma wave and statistically  produce
a net transfer of energy from the wave to the particles, the
plasmons  interact individually with the ripples and produce a
change of energy which is  more significant around the resonance,
which is determined by the condition $p=\omega /k$.  Finally, we
observe that since the system can be modulationally unstable
\cite{Zakharov}, this phenomenon acts in competition with the
modulational instability.

\end{document}